\documentclass[prl,twocolumn,preprintnumbers,amsmath,amssymb,superscriptaddress]{revtex4}

\usepackage[dvips]{graphicx}
\usepackage{dcolumn}
\usepackage{bm}
\usepackage{longtable}
\usepackage[dvips]{color}
\usepackage{ulem}

\begin{document}

\preprint{APS/123-QED}

\title{Deformation of the Magnetic Skyrmion Lattice in MnSi under Electric Current Flow}

\author{D. Okuyama}%
\email[]{okudaisu@tohoku.ac.jp}
\affiliation{Institute of Multidisciplinary Research for Advanced Materials (IMRAM), Tohoku University, Katahira 2-1-1, 
Sendai 980-8577, Japan.}
\author{M. Bleuel}%
\affiliation{NIST Center for Neutron Research, National Institute of Standards and Technology, 100 Bureau Drive, Gaithersburg, 
Maryland 20899-8562, USA}
\affiliation{Department of Materials Science and Engineering, University of Maryland, College Park, MD 20742-2115, USA}
\author{J. S. White}%
\affiliation{Laboratory for Quantum Magnetism, Institute of Physics, \'{E}cole Polytechnique F$\acute{e}$d$\acute{e}$rale de Lausanne (EPFL), 
CH-1015 Lausanne, Switzerland}
\affiliation{Laboratory for Neutron Scattering and Imaging (LNS), Paul Scherrer Institut (PSI), CH-5232 Villigen, Switzerland}
\author{Q. Ye}%
\affiliation{NIST Center for Neutron Research, National Institute of Standards and Technology, 100 Bureau Drive, Gaithersburg, 
Maryland 20899-8562, USA}
\affiliation{Department of Materials Science and Engineering, University of Maryland, College Park, MD 20742-2115, USA}
\author{J. Krzywon}%
\affiliation{NIST Center for Neutron Research, National Institute of Standards and Technology, 100 Bureau Drive, Gaithersburg, 
Maryland 20899-8562, USA}
\author{G. Nagy}
\affiliation{Laboratory for Neutron Scattering and Imaging (LNS), Paul Scherrer Institut (PSI), CH-5232 Villigen, Switzerland}
\author{Z. Q. Im}%
\affiliation{Laboratory for Quantum Magnetism, Institute of Physics, \'{E}cole Polytechnique F$\acute{e}$d$\acute{e}$rale de Lausanne (EPFL), 
CH-1015 Lausanne, Switzerland}
\author{I. \v{Z}ivkovi\'{c}}%
\affiliation{Laboratory for Quantum Magnetism, Institute of Physics, \'{E}cole Polytechnique F$\acute{e}$d$\acute{e}$rale de Lausanne (EPFL), 
CH-1015 Lausanne, Switzerland}
\author{M. Bartkowiak}%
\affiliation{Laboratory for Scientific Developments and Novel Materials (LDM), Paul Scherrer Institut (PSI), CH-5232 Villigen, Switzerland}
\author{H. M. R$\o$nnow}%
\affiliation{Laboratory for Quantum Magnetism, Institute of Physics, \'{E}cole Polytechnique F$\acute{e}$d$\acute{e}$rale de Lausanne (EPFL), 
CH-1015 Lausanne, Switzerland}
\author{S. Hoshino}%
\affiliation{RIKEN Center for Emergent Matter Science (CEMS), Wako 351-0198, Japan}
\author{J. Iwasaki}%
\affiliation{Department of Applied Physics, University of Tokyo, Tokyo 113-8656, Japan}
\author{N. Nagaosa}%
\affiliation{RIKEN Center for Emergent Matter Science (CEMS), Wako 351-0198, Japan}
\affiliation{Department of Applied Physics, University of Tokyo, Tokyo 113-8656, Japan}
\author{A. Kikkawa}%
\affiliation{RIKEN Center for Emergent Matter Science (CEMS), Wako 351-0198, Japan}
\author{Y. Taguchi}%
\affiliation{RIKEN Center for Emergent Matter Science (CEMS), Wako 351-0198, Japan}
\author{Y. Tokura}%
\affiliation{RIKEN Center for Emergent Matter Science (CEMS), Wako 351-0198, Japan}
\affiliation{Department of Applied Physics, University of Tokyo, Tokyo 113-8656, Japan}
\author{D. Higashi}%
\affiliation{Institute of Multidisciplinary Research for Advanced Materials (IMRAM), Tohoku University, Katahira 2-1-1, 
Sendai 980-8577, Japan.}
\author{J. D. Reim}%
\affiliation{Institute of Multidisciplinary Research for Advanced Materials (IMRAM), Tohoku University, Katahira 2-1-1, 
Sendai 980-8577, Japan.}
\author{Y. Nambu}%
\affiliation{Institute of Multidisciplinary Research for Advanced Materials (IMRAM), Tohoku University, Katahira 2-1-1, 
Sendai 980-8577, Japan.}
\affiliation{Institute for Materials Research (IMR), Tohoku University, Katahira 2-1-1, Sendai 980-8577, Japan.}
\author{T. J. Sato}%
\affiliation{Institute of Multidisciplinary Research for Advanced Materials (IMRAM), Tohoku University, Katahira 2-1-1, 
Sendai 980-8577, Japan.}

\date{\today}

\begin{abstract}
Using small-angle neutron scattering (SANS), we investigate the deformation of the magnetic skyrmion lattice 
in bulk single-crystalline MnSi under electric current flow.  
A significant broadening of the skyrmion-lattice-reflection peaks was observed in the SANS pattern for current densities 
greater than a threshold value $j_{\mathrm{t}}$ $\sim$~1~MA/m$^2$ (10$^6$ A/m$^2$).  
We show this peak broadening to originate from a spatially inhomogeneous rotation of the skyrmion lattice, 
with an inverse rotation sense observed for opposite sample edges aligned with the direction of current flow.  
The peak broadening (and the corresponding skyrmion lattice rotations) remain finite even after switching off the electric current.  
These results indicate that skyrmion lattices under current flow experience significant friction near the sample edges, 
and plastic deformation due to pinning effects, 
these being important factors that must be considered for the anticipated skyrmion-based applications in chiral magnets at the nanoscale.  
\end{abstract}
                              
\maketitle
A magnetic skyrmion is a swirling spin texture characterized by a discrete topological number, called the skyrmion number.  
Such topologically nontrivial states of matter were first theoretically predicted in nonlinear field theory~\cite{Skyrme1961,Derrick1964}, 
and subsequently in chiral magnets~\cite{Bogdanov1989}.  
Experimentally, magnetic skyrmions often condense into triangular-lattice, observed as six-fold magnetic Bragg reflections 
in small-angle neutron scattering (SANS), first discovered 
by M$\ddot{\mathrm{u}}$hlbauer {\it et al.} in the prototypical chiral magnet MnSi~\cite{Mublbauer2009}.  
To date, triangular skyrmion-lattice structures are widely confirmed in various magnets ranging from metallic to insulating compounds~\cite{Nagaosa2013,Neubauer2009,Bauer2012,Seki2012a,Adams2012,Seki2012b}, 
and also by various techniques such as Lorentz transmission electron microscopy (TEM) 
and magnetic force microscopy~\cite{Yu2010,Milde2013}.

There are several prominent characteristics of magnetic skyrmions that make them quite intriguing for fundamental 
as well as technological application viewpoints.  
One is its topological protection; once created, the skyrmion can hardly be annihilated~\cite{Nagaosa2013}.  
Hence, there is a possibility for the magnetic skyrmion to be used as a robust information storage/carriage unit in future spintronics devices.  
In metallic skyrmion compounds, there is another important characteristic, namely its strong coupling to electric current flow.  
The electric current density required to realize the skyrmion-lattice motion is remarkably small at $j \sim$ 1~MA/m$^2$ (10$^6$ A/m$^2$), 
compared to that required to move magnetic domain boundaries in conventional magnetic materials 
$j \geq$ 1~GA/m$^2$ (10$^9$ A/m$^2$)~\cite{Myers1999,Grollier2003,Tsoi2003,Yamanouchi2004}.  
The potential for controlling magnetic skyrmions with such a low current density, together with their topologically protected nature, 
makes spintronics application promising~\cite{Fert2013}.  
Hence, magnetic skyrmions continue to attract strong attention, and remain under intense scrutiny for elucidating their dynamical behavior 
under electric current.  

A number of reports have appeared to date on the motion of magnetic skyrmions under electric current flow.  
In the pioneering study of Jonietz {\it et al.} using SANS, a rotation of the skyrmion lattice in MnSi was detected 
for an electric current density greater than the threshold value $j_{\mathrm{t}} \sim$ 1~MA/m$^2$~\cite{Jonietz2010}.  
There, it was argued that the electric current flow alone introduces translational motion only via spin transfer torques, 
which cannot be detected by SANS.  
Hence, a thermal gradient was intentionally applied in their experiment, which provides spatial inhomogeneity of the spin transfer torque.  
Indeed, the skyrmion-lattice rotation was observed only when both the current flow and thermal gradient were present.  
Nonetheless, by careful examination of the presented data, it can be seen that the skyrmion-lattice peaks are significantly broadened 
under the electric current flow, the origin of which was not discussed.  

Skyrmion-lattice motion under thermally homogeneous conditions was studied in the other works.  
From Hall-resistivity measurements on MnSi, the skyrmions move in the direction of the electric current 
above a critical current density $j_{\mathrm{t}} \sim$ 0.5~MA/m$^2$, 
while the emergent electric field is transverse to the current~\cite{Schulz2012,Zang2011}.  
From a Lorentz TEM study of FeGe, the skyrmion lattice was observed to \textquoteleft disappear\textquoteright\ 
as the electric current exceeded a threshold value, indicating the skyrmion lattice to move much faster than 
the Lorentz TEM time frame~\cite{Yu2012}.  
These studies show that the electric current flow induces the skyrmion-lattice motion.  
However, important microscopic information, such as the skyrmion-lattice deformation, 
under thermally homogeneous conditions remains largely unexplored.  

Here, we investigate the deformation of the magnetic skyrmion lattice under an electric current flow using the SANS technique, 
paying careful attention towards keeping any thermal gradient as small as experimentally achievable.  
We observe clearly a broadening of the skyrmion-lattice-reflection peaks for $j > j_{\mathrm{t}}$, 
indicating the skyrmion lattice to deform considerably when the lattice starts to flow translationally.  
We further find that the peak broadening is due to a spatially inhomogeneous rotation of the skyrmion lattice; 
by measuring opposing-sample edges, we observed a counter-rotating behavior of the skyrmion lattice.  
While the rotation direction does not change under inversion of the external magnetic field, 
it does revert when the electric current direction is reversed.  
Finally, the peak broadening and corresponding lattice rotation remain finite even after the current is turned off, 
indicating a plastic deformation of the skyrmion lattice under the force exerted by the electric current.  

\begin{figure}
\includegraphics*[width=80mm,clip]{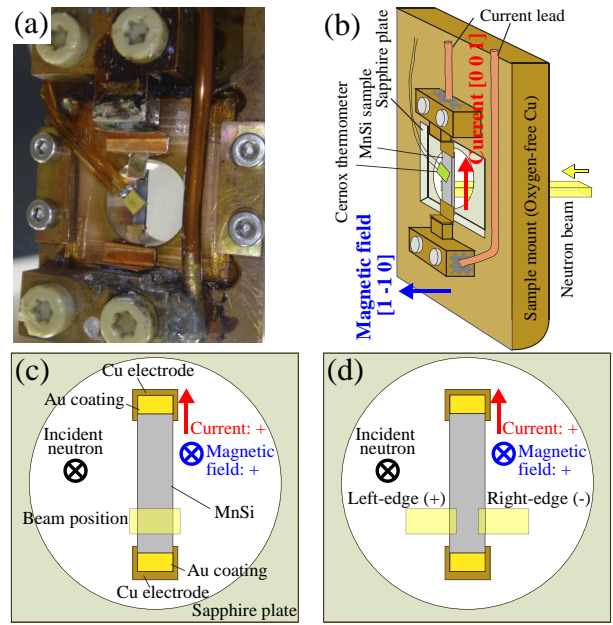}
\caption{\label{fig_01} 
Photograph (a) and schematic drawing (b) of the MnSi sample and sample mount for the NG7 experiment.  
(c, d) Schematic illustration of the beam illumination area for investigating the peak broadening (c) 
and counter-rotating behavior at the sample edges (d) of the skyrmion-lattice peaks.  
}
\end{figure}

SANS experiments were carried out at NG7 (National Institute of Standards and Technology) and SANS-II (Paul Scherrer Institut).  
The single-crystal-MnSi samples were cut in a rectangular shape of 1.4 mm (width) $\times$ 7.5 mm (height) 
$\times$ 0.4 mm (thickness) for NG7.  
The incident neutron wavelength was $\lambda_{\rm i}$ = 6~\AA\ with $\rm \Delta \lambda / \lambda_{\rm i}$ = 14 \%.  
To further limit any temperature inhomogeneity in the measured region, and also to check the sample position dependence, 
only a small part of the sample was illuminated by using a very narrow beam of 2.0 mm (width) $\times$ 1.0 mm (height).  
Full details of our experiments can be found in the Supplemental Material~\cite{Supple2018}.  

A photograph of the sample mount for the NG7 experiment and its schematic illustration are shown in Figs.~\ref{fig_01}(a) 
and \ref{fig_01}(b), respectively.  
An electric current up to 2~A ($j$ = 3.6~MA/m$^{2}$) was applied along the [0 0 1] direction~\cite{thermometer}.  
The temperature gradient along the current-flow direction was estimated by measuring the position dependence 
of the ordering temperature, and was confirmed to be less than 0.035~K/mm at the sample region investigated 
in the present study under $j$ = 2.7~MA/m$^2$~\cite{Supple2018}.  
This is at least one order of magnitude smaller than in the previous experiment~\cite{Jonietz2010}.  
The sample mount was installed into a horizontal-field magnet with the magnetic field applied 
along [1 -1 0] parallel to the incident neutron beam.  

\begin{figure}
\includegraphics*[width=80mm,clip]{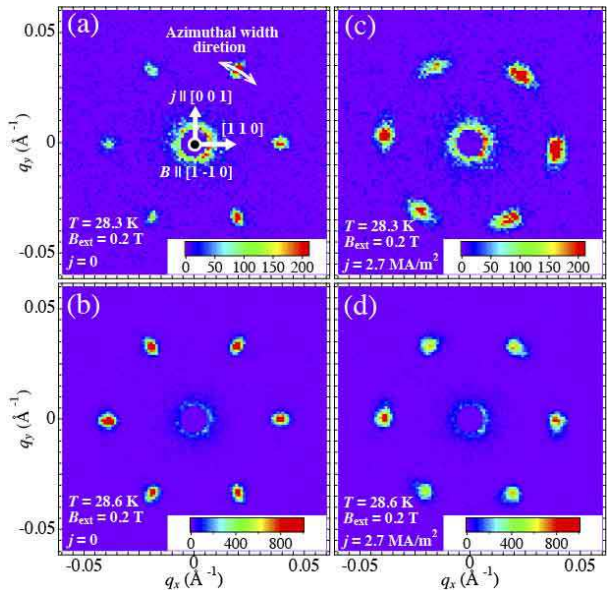}
\caption{\label{fig_02} 
(a)-(d) SANS measured at $B_{\mathrm{ext}}$ = 0.2~T.  
The skyrmion reflections were measured at $T$ = 28.3~K (28.6~K) under $j$ = 0 (2.7~MA/m$^2$), respectively.  
All data were measured for 10 minutes.  
When the intensities of the six skyrmion reflections are not equivalent, 
the sample is slightly misoriented away from the Bragg position.  
The azimuthal width direction is indicated by the arrow in panel (a).  
}
\end{figure}

Representative SANS patterns measured at NG7 are shown in Figs.~\ref{fig_02}(a)-\ref{fig_02}(d).  
They were measured from imaging an entire horizontal cross-section of the sample, as shown in Fig.~\ref{fig_01}(c).  
Figures \ref{fig_02}(a) and \ref{fig_02}(b) show the skyrmion reflections observed at the two temperatures, $T$ = 28.3~K and 28.6~K, respectively, under $B_{\mathrm{ext}}$ = 0.2~T and $j$ = 0~\cite{magfield}.  
The six-fold magnetic reflections characteristic of the skyrmion lattice were observed 
in a temperature range of 28~K $< T <$ 29.2~K at $B_{\mathrm{ext}}$ = 0.2~T~\cite{Supple2018}.  
These observations are identical to those in the earlier works~\cite{Mublbauer2009,Jonietz2010}.  

Figures \ref{fig_02}(c) and \ref{fig_02}(d) show the SANS patterns obtained at $j = 2.7$~MA/m$^2$ and $B_{\mathrm{ext}}$ = 0.2~T.  
By comparing with the zero-current-flow counterparts in Figs.~\ref{fig_02}(a) and \ref{fig_02}(b), 
a considerable azimuthal broadening of the skyrmion-lattice peaks was observed.  
The peak broadening is temperature dependent at $j = 2.7$~MA/m$^2$; 
the width is considerably larger at $T = 28.3$~K compared to that at $T = 28.6$~K.  
This result clearly indicates the significant deformation of the skyrmion lattice under large electric current.  

\begin{figure}
\includegraphics*[width=80mm,clip]{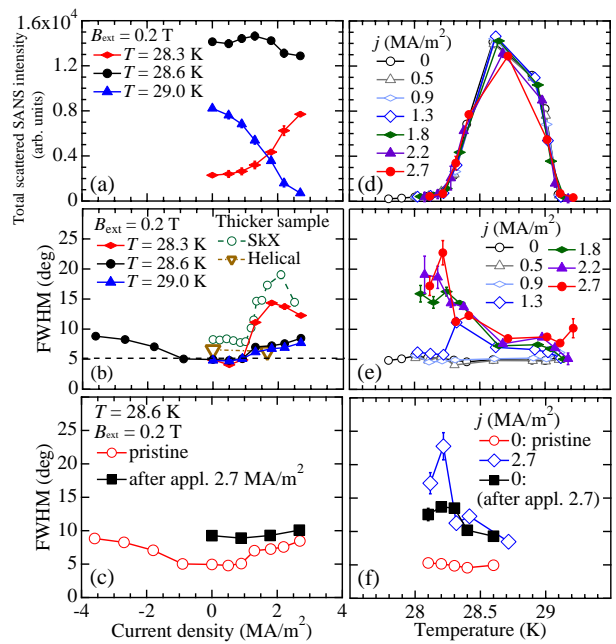}
\caption{\label{fig_03} 
Current-density dependences of (a) the total scattered SANS intensity and (b), (c), the azimuthal spot widths of the skyrmion reflections 
measured at three temperatures.  
The temperature used in (a)-(c) is the sensor temperature, and therefore not the true temperature estimated from the ordering temperature 
of the helical phase.  
Thus, the increase/decrease of the intensity at $T$ = 28.3~K/29.0~K by applying the electric current is due to sample heating.  
In (b), the azimuthal width of the skyrmion ($B_{\mathrm{ext}}$ = 0.2 T) and helical ($B_{\mathrm{ext}}$ = 0) reflections at $T$ = 28.6 K 
measured by the thicker sample are also shown.  
Temperature dependences of (d) the total scattered SANS intensity and (e, f) azimuthal widths of the skyrmion reflections 
at each current density.  
In (c) and (f), additionally the dependence of the width after applying $j$ = 2.7~MA/m$^2$ is shown.  
}
\end{figure}

To examine quantitatively the current dependence of the skyrmion-lattice peaks, in Figs. \ref{fig_03}(a) and \ref{fig_03}(b), 
we show respectively the total scattered SANS intensity and azimuthal spot widths as functions of current density 
at three selected temperatures and $B_{\mathrm{ext}}$ = 0.2 T.  
The spot width remains almost constant in the low current density region ($j < j_{\mathrm{t}} \sim 1$~MA/m$^2$) for all measured temperatures.  
At $T$ = 28.3~K, the width shows a steep increase for $j > j_{\mathrm{t}}$.  
This clearly confirms the existence of a threshold current density for skyrmion-lattice deformation.  
It may be noted that the value of $j_{\mathrm{t}}$ is in good agreement with the current density above which skyrmion-lattice motion 
was detected in the earlier studies~\cite{Jonietz2010,Schulz2012}.  
Hence, the peak broadening detected above $j_{\mathrm{t}}$ in the present study is a direct consequence of 
skyrmion-lattice motion driven by the electric current.  
In stark contrast to the skyrmion-lattice peaks, we observed no peak broadening for the reflections in the helical phase 
at $B_{\mathrm{ext}}$ = 0 up to 1.67~MA/m$^2$, as shown by the open triangular symbol in Fig.~\ref{fig_03}(b).  

Once induced by the electric current, the broadening of the skyrmion-lattice peaks persists, even after the electric current is removed.  
In Fig.~\ref{fig_03}(c), starting from the pristine state, the width increases monotonically up to $j = 2.7$~MA/m$^2$ above $j_{\mathrm{t}}$, 
whereas when lowering the current, the width stays almost the same down to $j = 0$.  
This clearly indicates that the skyrmion lattice remembers the deformation generated in the driven state.  
In other words, the skyrmion lattice locally deforms like plastic matter when driven under electric current.  
This plastic deformation may be due to impurity pinning~\cite{plasticity}.  

Figures \ref{fig_03}(d) and \ref{fig_03}(e) show the temperature dependences of respectively the total scattered SANS intensity 
and azimuthal spot widths at fixed current density.  
At $j$ = 2.7~MA/m$^2$, the width shows a significant temperature dependence; 
it monotonically increases as temperature is decreased.  
Below $j_{\mathrm{t}}$, the width shows negligible temperature dependence.  
Figure \ref{fig_03}(f) shows the temperature dependence of the width at $j$ = 0, measured with decreasing temperature, 
after having applied the current density 2.7~MA/m$^2$ at $T$ = 28.6~K.  
Independent of the temperature, the spot widths remain broadened compared with the pristine state at $j$ = 0 
after the current ramping, confirming the robust memory effect for the skyrmion-lattice deformation.  

\begin{figure}
\includegraphics*[width=80mm,clip]{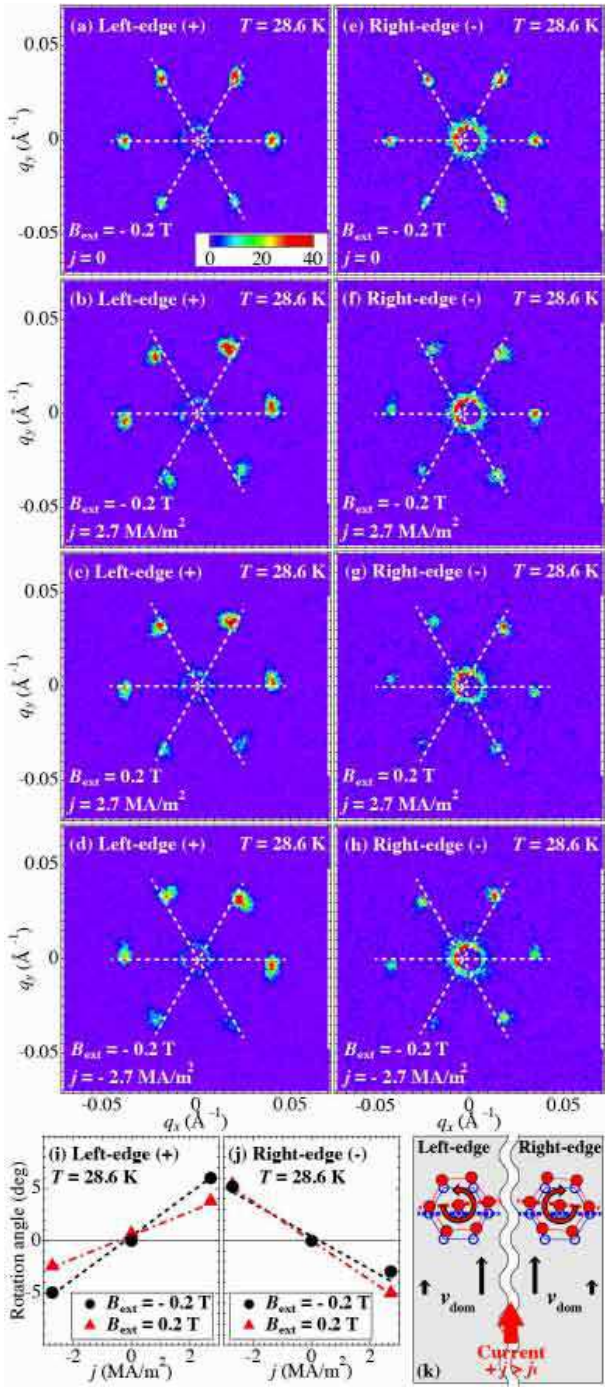}
\caption{\label{fig_04} 
SANS measured from the left-edge (+) or right-edge (-) of the sample as schematically illustrated in Fig. \ref{fig_01}(d).  
(a)-(h) The skyrmion reflections at $T$ = 28.6~K respectively measured at $B_{\mathrm{ext}}$ = -0.2 (0.2~T), 
$j$ = 0 (2.7~MA/m$^2$), and at the left-edge (right-edge).  
All data were measured for 20~minutes.  
Dashed lines are guides to eye for the peak positions of the skyrmion reflections in pristine condition.  
The current-density and magnetic-field dependences of the averaged rotation angles of the skyrmion reflections 
at $T$ = 28.6~K at (i) left- and (j) right-edges.  
The circle and triangle denote the data at $B_{\mathrm{ext}}$ = -0.2 and 0.2~T, respectively.  
The dashed lines are guides to eye.  
(k) Schematic illustration of the skyrmion-lattice rotation above $j_{\mathrm{t}}$.  
Blue and red circles stand for the pristine and moved skyrmions, respectively.  
Black arrows and $v_{\mathrm{dom}}$ explain the velocity of the skyrmion-lattice domain near the sample edges.  
}
\end{figure}

To investigate the origin of the broadening of the skyrmion reflections, we performed SANS measurements on the left-edge (+) or right-edge (-) 
of the sample, as shown schematically in Fig. \ref{fig_01}(d).  
The size of the neutron illumination area is approximately 0.2 mm (width) $\times$ 1.0 mm (height).  
The SANS patterns obtained at $T$ = 28.6~K and $B_{\mathrm{ext}}$ = -0.2~T without electric current are shown in Figs.~\ref{fig_04}(a) 
and \ref{fig_04}(e) for the left- and right-edges, respectively.  
The observed patterns are identical for the two edges, and in good agreement with the data shown in Fig.~\ref{fig_02}(b).  
In marked contrast to the zero-current condition, under $j$ = 2.7~MA/m$^2$ the reflection patterns taken from the left- and right-edge parts 
exhibit counterclockwise and clockwise rotations, respectively, as shown in Figs.~\ref{fig_04}(b) and \ref{fig_04}(f).  

Next, the effect of inverting the current and magnetic-field directions was investigated.  
Figures~\ref{fig_04}(c) and \ref{fig_04}(g) show data taken under an inverted magnetic field $B_{\mathrm{ext}}$ = 0.2~T.  
Apparently, the rotation direction does not change by inverting $B_{\mathrm{ext}}$.  
On the other hand, the inversion of the current direction (from $j = +2.7$ to $-2.7$~MA/m$^2$) results in a sign change of the rotation angle, 
as clearly seen by comparing Figs.~\ref{fig_04}(d) and \ref{fig_04}(h) ($j = -2.7$~MA/m$^2$) 
with Figs.~\ref{fig_04}(b) and \ref{fig_04}(f) ($j = +2.7$~MA/m$^2$).  
This indicates that the skyrmion-lattice rotation observed in the thermally homogeneous condition depends not on the magnetization direction, 
but only on the current direction.  
It should be noted that the transverse translational motion is due to the scalar spin chirality, which depends 
on the third order of the magnetization density $\vec{M}(\vec{R})$, whereas the longitudinal translational mode along the electric current is 
due to the dissipative tensor which is in the second order in $\vec{M}(\vec{R})$~\cite{Everschor2011,Thiele1972,Gyro}.  
The independence of the observed rotation angle to the magnetization direction thus indicates that the rotation effect should be attributed to 
the longitudinal translational motion of the skyrmions under the electric current.  
The averaged rotation angles for the six skyrmion reflections are plotted in Figs.~\ref{fig_04}(i) and \ref{fig_04}(j) 
for the left- and right-edge parts, respectively.  
A counter-rotating behavior that is clearly antisymmetric around $j$ = 0 is observed for the two edge positions 
with a rotation angle of approximately $\pm 5$~degrees for $j = 2.7$~MA/m$^2$ at $T = 28.6$~K and $B_{\mathrm{ext}}$ = $\pm 0.2$~T.  
Hence, we conclude that the peak broadening is mainly due to a spatially inhomogeneous rotation of the skyrmion lattice, 
with counter-rotating behavior for the left- and right-edges.  
It should be further noted that the sum of absolute rotation angle at each edge is in semiquantitative agreement with 
the overall broadening observed from imaging across the whole sample [Fig. \ref{fig_01}(c)], supporting our conclusion.  

Plausibly related to the present observation is a recent magneto-optical Kerr study of the motion of {\it single} skyrmions under current flow 
in a CoFeB-based trilayer, where the velocity of the skyrmion near the edge of the sample is reported to be 40~\% less 
than that at the center of the sample~\cite{Jiang2017}.  
It has also been found that the pinning effect is stronger when the confining potential is acting on the skyrmions 
near the edge~\cite{Iwasaki2014}.  
Therefore, we expect that a shear flow of the skyrmion-lattice domains may appear in our sample where 
the velocity of the skyrmion-lattice domains is higher in the middle region and slower by the pinned motion near the edges.  
Such a shear flow generates a rotational torque on the skyrmion-lattice domains as shown in Fig. \ref{fig_04}(k), 
which may be the origin of the counter-rotating behavior observed in the present study.  
In other words, the skyrmion lattice flows with a shear component plausibly due to significant friction near the edges.  
From the rotation direction of the skyrmion lattice relative to the direction of the electric current as shown in Figs.~\ref{fig_04}(b)-\ref{fig_04}(g), 
we can deduce that the skyrmion motion in MnSi is along the charge current not the electron current.  
This is in agreement with the conclusion based on the topological Hall resistivity measurement that macroscopically 
the skyrmions in MnSi drift in the same direction as the electric current flows~\cite{Schulz2012,privatecomm_01}.  

In the simple particle model, where skyrmions are dense and move randomly, this edge effect works only 
near the edge ($\sim$ 1 nm).  
To explain the generation of a shear stress deeper inside the sample, a numerical simulation including inter-skyrmion interactions 
with taking into account friction amongst the skyrmions as well as the robustness of their formed triangular lattice, is required.  

Lastly, we comment on the effect of the thermal gradient.  
Despite our effort to minimize the thermal gradient, a small but finite gradient remained along the electric-current direction.  
This thermal gradient was pointed out as the origin of the skyrmion lattice rotation in Ref.~\cite{Jonietz2010} and~\cite{Everschor2012}.  
However, since the rotational motion under the thermal gradient is due to the spin transfer torque, which should 
depend on the macroscopic magnetization direction, 
one expects that the sign of the rotation angle is switched when the external magnetic field is reversed.  
We clearly found that the rotation angle at each edge does not change upon inversion of the magnetic-field direction, 
which excludes the possibility of the remaining thermal gradient giving rise to the skyrmion-lattice rotation in the present setup.  

In summary, we have used SANS to study skyrmion-lattice motion in bulk MnSi under electric current flow in a thermally homogenous condition.  
The azimuthal width of the skyrmion-lattice peaks shows significant broadening above a threshold current density 
$j_{\mathrm{t}} \sim$~1~MA/m$^{2}$.  
We show this peak broadening originates from a spatially inhomogeneous rotation of the skyrmion lattice, that shows 
opposite senses of rotation at the sample edges.  
An intriguing memory effect was also observed for the peak broadening and the corresponding rotation of the skyrmion lattice, 
indicating that the driven skyrmion lattice considerably deforms due to the friction near the sample edges.   

The authors thank P. D. Butler, K. Everschor, M. Garst, A. Rosch, T. Hanaguri, W. Koshibae, J. Barker, 
A.P. Tsai, K. Yamauchi and T. Oguchi for the fruitful discussions.  
This work was in part supported by Grants-in-Aids for Scientific Research (No. 24224009, 26103006, 17K14327, 17K18744, and 18H03676) 
from the Ministry of Education, Culture, Sports, Science and Technology (MEXT), Japan 
and by the Research Program for CORE lab of \textquotedblleft Dynamic Alliance for Open Innovation Bridging Human, 
Environment and Materials\textquotedblright\ in \textquotedblleft Network Joint Research Center for Materials and Devices\textquotedblright.  
The Swiss National Science Foundation (SNF) Sinergia network \textquotedblleft NanoSkyrmionics\textquotedblright\ 
(Grant No. CRSII5-171003), the SNF projects 153451 and 166298, and the European Research Council project CONQUEST.


\begin{thebibliography}{200}
\bibitem{thermometer}
For such a large electric current, the sample temperature deviates from the sensor temperature slightly due to self-heating.  
Hence, we estimated the sample temperature using the ordering temperature of the zero-field helical phase; 
the maximum difference between the sensor and sample temperatures was 0.16~K at $j$ = 2.7~MA/m$^2$.
\bibitem{plasticity}
From the peak width of the skyrmion reflection, the domain size is small with less than 10 skyrmion unit cells.  
It is expected that the skyrmion-lattice domain deforms with dislocation slipping along the domain boundary.  
This kind of deformation is called plastic deformation.  
Thus, in this paper, we use the terminology of \textquoteleft plastic deformation\textquoteright\ to explain the observed shear distortion 
of the skyrmion lattice.
\end{thebibliography}
\end{document}